\documentclass[conf]{new-aiaa}
\usepackage[utf8]{inputenc}
\usepackage{subdepth}
\usepackage{graphicx}
\usepackage{verbatim}
\usepackage{amsmath}
\usepackage[version=4]{mhchem}
\usepackage{siunitx}
\usepackage{longtable,tabularx}
\usepackage{multicol}
\usepackage{subcaption} 
\usepackage{algorithm}
\usepackage{algpseudocode}
\usepackage{todonotes}
\usepackage{pythonhighlight}

\usepackage{tcolorbox}
\definecolor{ucsd_blue}{RGB}{24, 43, 73}
\definecolor{ucsd_gold}{RGB}{198, 146, 20}
\definecolor{ucsd_blue_light}{RGB}{0, 98, 155}
\definecolor{ucsd_gold_light}{RGB}{255, 205, 0}
\definecolor{ucsd_gray}{RGB}{116, 118, 120}

\title{Graph-accelerated non-intrusive polynomial chaos expansion using partially tensor-structured quadrature rules for uncertainty quantification}

\author{Bingran Wang
, Nicholas C. Orndorff and John T. Hwang}
\affil{University of California, San Diego, La Jolla, CA, USA}

\begin{document}

{\begin{tcolorbox}[boxrule=0.75pt, arc=2pt, coltext=ucsd_blue, colback=white,colframe=ucsd_gold, fontupper=\rmfamily]
        \footnotesize
        \small
        \setstretch{1.25}
        {\textcolor{ucsd_gray}
        {This is the preprint version of the following article:}}
        
        Bingran Wang, Nicholas C. Orndorff, and John T. Hwang. Graph-accelerated non-intrusive polynomial chaos expansion using partially tensor-structured quadrature rules for uncertainty quantification, Aerospace Science and Technology, vol. 155, 2024, 109607.

        \vspace{1mm}
        \urlstyle{rm}
        \textcolor{ucsd_gray}{Published article:} \;
        \url{https://doi.org/10.1016/j.ast.2024.109607}

        \textcolor{ucsd_gray}
        {Preprint pdf:} \; 
        \url{https://github.com/LSDOlab/lsdo_bib/blob/main/pdf/wang2024partial.pdf}


        \textcolor{ucsd_gray}
        {Bibtex:} \;    \url{https://github.com/LSDOlab/lsdo_bib/blob/main/bib/wang2024partial.bib}

    \end{tcolorbox}
    
    \vspace{-6mm}
}

\maketitle
 \begin{abstract}
Recently, the graph-accelerated non-intrusive polynomial chaos (NIPC) method has been proposed for solving uncertainty quantification (UQ) problems. 
This method leverages the full-grid integration-based NIPC method to address UQ problems while employing the computational graph transformation approach, AMTC, to accelerate the tensor-grid evaluations.
This method exhibits remarkable efficacy on a broad range of low-dimensional (three dimensions or less) UQ problems featuring multidisciplinary models. 
However, it often does not scale well with problem dimensions due to the exponential increase in the number of quadrature points when using the full-grid quadrature rule.
To expand the applicability of this method to a broader range of UQ problems, this paper introduces a new framework for generating a tailored, partially tensor-structured quadrature rule to use with the graph-accelerated NIPC method. 
This quadrature rule, generated through the designed quadrature approach, possesses a tensor structure that is tailored for the computational model. 
The selection of the tensor structure is guided by an analysis of the computational graph, ensuring that the quadrature rule effectively capitalizes on the sparsity within the computational graph when paired with the AMTC method.
This method has been tested on one 4D and one 6D UQ problem, both originating from aircraft design scenarios and featuring multidisciplinary models.
Numerical results show that, when using with graph-accelerated NIPC method, our approach generates a partially tensor-structured quadrature rule that outperforms the full-grid Gauss quadrature and the designed quadrature methods (more than 40\% reduction in computational costs) in both of the test problems.
\end{abstract}
\section{Introduction}
Engineers rely heavily on computational models to design, analyze, and optimize engineering systems. However, such computational models are often deterministic and inevitably affected by uncertainties inherent in real-world scenarios. Uncertainty quantification (UQ) plays a pivotal role in ensuring the reliability of predictions made by these computational models. Its applications span various scientific and engineering domains, including weather forecasting~\cite{joslyn2010communicating,hess-9-381-2005}, machine learning~\cite{hullermeier2021aleatoric, abdar2021review}), structural analysis~\cite{wan2014analytical,hu2018uncertainty}, and aircraft design~\cite{ng2016monte,wang2024graph,lim2022uncertainty}, where uncertainties can significantly impact system behavior and performance.

Forward UQ, also known as uncertainty propagation, seeks to assess how uncertainties associated with model inputs affect a computational model's outputs or quantities of interest (QoIs). 
These input uncertainties can stem from variations in operating conditions and model parameters, referred to as aleatoric uncertainties, or from model errors due to incomplete knowledge, termed epistemic uncertainties. By quantifying the impact of input uncertainties, forward UQ plays a crucial role in supporting decision-making and risk assessment in engineering system design.
In this paper, we address a UQ problem within a probabilistic framework, representing uncertain inputs as continuous probability density distributions. The objective is to estimate the statistical moments or risk measures of the QoI. 

We focus on employing the non-intrusive polynomial chaos (NIPC) method~\cite{xiu2002wiener} to solve the UQ problem, a popular approach especially suitable for UQ problems with fewer than 10 dimensions. 
The NIPC method approximates the model response through a linear combination of polynomial chaos expansion (PCE) basis functions, determined by the distributions of the random inputs. 
The PCE coefficients are computed using either sampling methods~\cite{jones2013nonlinear,hosder2007efficient,blatman2011adaptive} or integration methods~\cite{keshavarzzadeh2017topology, xiaojing2018sparse, luo2021robust}. 
In \cite{eldred2009comparison}, Eldred and Burkardt demonstrated that for low-dimensional UQ problems, the integration-based and regression-based NIPC methods show comparable performance, with significantly higher convergence rates than the Monte Carlo method.

The integration approach makes use of the orthogonality properties of the PCE basis functions and transforms the UQ problems into multidimensional integration problems. Common methods for solving the integration problems include Gauss quadrature~\cite{golub1969calculation}, Smolyak sparse-grid~\cite{smolyak1963quadrature}, and designed quadrature~\cite{keshavarzzadeh2018numerical} methods.
The Gauss quadrature method formulates the quadrature rule using a tensor product of univariate quadrature rules. This approach is easy to use but the number of quadrature points increases exponentially with the number of dimensions.
The Smolyak sparse-grid approach breaks the tensor structure of the Gauss quadrature rule by dropping the higher-order cross terms while achieving minimal sacrifice on its accuracy.
In comparison, the design quadrature method generates the quadrature rule by solving an optimization problem to find approximate solutions to the moment-matching equations.
The designed quadrature method results in a quadrature rule that does not possess a tensorial structure but can typically achieve better performance than the previous two methods.

Wang et al. recently introduced a new method called \textit{Accelerated Model evaluations on Tensor grids using Computational graph transformations} (AMTC)~\cite{wang2023accelerating}. 
This method reduces the computational cost of model evaluations on tensor-grid inputs by modifying the computational graph of the model within the middle end of a modeling language's three-stage compiler. This modification eliminates redundant evaluations at the operation level caused by the tensor structure of the inputs.
Their previous work applies AMTC with integration-based NIPC to solve UQ problems, under the assumption of a specific type of sparsity within the computational graphs of the computational models. 
This integration is termed the \textit{graph-accelerated NIPC} method and has demonstrated superior efficiency in various UQ and optimization under uncertainty problems involving multidisciplinary systems~\cite{wang2024graph}.
However, the graph-accelerated NIPC method has so far only been implemented with a full-grid Gauss quadrature rule, targeting low-dimensional UQ problems.

The primary contribution of our paper is a new framework to generate partially tensor-structured quadrature rules that can be used with the graph-accelerated NIPC methods to solve UQ problems with multidisciplinary models. The proposed method generates the quadrature rule following using a particular tensor structure such that the associated model evaluations can be accelerated using the recently proposed computational graph acceleration method, AMTC.
Within this framework, we first identify a suitable tensor structure option for the quadrature rule through an analysis of the computational model. Subsequently, we employ the designed quadrature method to construct the quadrature rule that adheres to this tensor structure.
The resulting quadrature rule is strategically designed to leverage the sparsity inherent in the computational graph of the model, thus maximizing efficiency when employing the AMTC method for accelerating tensor-grid evaluations.

This method has been applied to two UQ problems, one with four uncertain inputs and one with six uncertain inputs, both derived from practical aircraft design problems involving multidisciplinary systems.
The proposed method produces new partially tensor-structured quadrature rules distinct from existing ones for both test problems. 
These partially structured quadrature rules also outperform existing methods when employed with the graph-accelerated NIPC method for solving UQ problems.

This paper is structured as follows. Section \ref{Sec: Background} provides background information on NIPC, quadrature rules, and AMTC. 
In Section \ref{Sec: Methodology}, we elaborate on the proposed framework. Section \ref{Sec: Numerical Results} presents the numerical results obtained from the test problems. 
Finally, Section \ref{Sec: Conclusion} summarizes the work and provides concluding remarks.

\section{Background}
\label{Sec: Background}
\subsection{Non-intrusive polynomial chaos}
Polynomial chaos, also known as polynomial chaos expansion (PCE), traces its origins back to Wiener, who utilized Hermite polynomials in developing a theoretical framework for Gaussian random processes~\cite{wiener1938homogeneous}. 
Ghanem further expanded upon this concept, employing Hermite polynomials as orthogonal bases to represent random processes~\cite{ghanem1999ingredients}. 
However, this method suffers from convergence issues for problems with non-Gaussian uncertain inputs.  
This issue was addressed by Xiu and Karniadakis in their work~\cite{xiu2002wiener}, wherein they introduced the generalized polynomial chaos (gPC) method. 
The gPC method strategically employs different orthogonal polynomials tailored to uncertain inputs with different distributions, thereby achieving optimal convergence in addressing UQ problems.

We consider a UQ problem involving a function represented as
\begin{equation}
f = \mathcal{F}(u),
\end{equation}
where $\mathcal{F}:\mathbb{R}^{d} \to \mathbb{R}$ represents a deterministic model evaluation function, $u \in \mathbb{R}^d$ denotes the input vector, and $f \in \mathbb{R}$ represents a scalar output. 
The UQ problem aims to compute the statistical moments or risk measures of the stochastic $f(U)$ under the effect of the input uncertainties, denoted as a stochastic vector, $U := [U_1, \ldots, U_d]$. The input uncertainties are assumed to be mutually independent and adhere to the probability distribution $\rho(u)$ with support $\Gamma$.

Using the gPC method, the stochastic output can be represented as a linear combination of orthogonal polynomials:
\begin{equation}
\label{eqn: PCE}
    f(U) = \sum_{i = 0}^{\infty} \alpha_i \Phi_i(U),
\end{equation}
where $\Phi_i(U)$ denotes the PCE basis functions that are chosen based on the uncertain inputs' probability distribution $\rho(u)$, and $\alpha_i$ represents the PCE coefficients that need to be determined.
In practical applications, it is common to truncate the PCE basis functions up to a specific polynomial order and approximate the stochastic output as
\begin{equation}
\label{Eqn: pce estimate}
    f(U) \approx \sum_{i = 0}^{q} \alpha_i \Phi_i(U).    
\end{equation}
With $p$ as the upper bound for the total polynomial order, the resultant number of PCE basis functions, $q+1$, satisfies
\begin{equation}
    q + 1 = \frac{(d + p)!}{d
    ! p!}.
\end{equation}
The PCE basis functions are chosen to satisfy the orthogonality property as
\begin{equation}
\label{Eqn: orthogonal property}
    \left< \Phi_i(U), \Phi_j(U) \right> = \delta_{ij},
\end{equation}
where $\delta_{ij}$ is the Kronecker delta and the inner product is defined as
\begin{equation}
    \left< \Phi_i(U), \Phi_j(U) \right> = \int_{\Gamma}  \Phi_i(u) \Phi_j(u)\rho(u) du.
\end{equation}
In the one-dimensional case, the PCE basis functions consist of univariate orthogonal polynomials chosen according to the distribution of the uncertain input~\cite{xiu2002wiener}.
Tab.~\ref{tab: orthog poly} presents common types of continuous random variables alongside their corresponding orthogonal polynomials and support ranges.
In multi-dimensional cases, the PCE basis functions are formed as tensor products of the univariate orthogonal polynomials. For instance, in the 2D case, the first six PCE basis functions are expressed as
\begin{equation*}
\begin{aligned}
    &\Phi_0(U)= \phi_0(U_1) \phi_0(U_2) \\
    &\Phi_1(U) = \phi_1(U_1) \phi_0(U_2) \\
    &\Phi_2(U) = \phi_0(U_1) \phi_1(U_2) \\
    &\Phi_3(U) = \phi_2(U_1) \phi_0(U_2) \\
    &\Phi_4(U) = \phi_1(U_1) \phi_1(U_2) \\
    &\Phi_5(U) = \phi_0(U_1) \phi_2(U_2), \\
\end{aligned}
\end{equation*}
where $\phi_i(U_j)$ represents the $i$-th univariate orthogonal polynomial corresponding to the uncertain input, $U_j$. If the uncertain inputs are dependent, the PCE basis functions can be generated numerically following the approach outlined in~\cite{lee2020practical}.

Using the gPC approximation in \ref{Eqn: pce estimate}, the statistical moments of the QoI can be analytically computed as
\begin{equation}
   \mu_f= \alpha_0,
\end{equation}
\begin{equation}
   \sigma_f = \sum_{i = i}^d\alpha_i^2 \left< \Phi_i^2\right>.
\end{equation}
For risk measures like probability of failure or Conditional Value at Risk (CVaR), the gPC model can be used as an inexpensive-to-evaluate surrogate model, and one can apply sampling methods to approximate them.
\begin{table}[]
\caption{Orthogonal polynomials for common types of continuous random variables}
\centering
\begin{tabular}{c c c c} 
 \hline
 Distribution & Orthogonal polynomials & Support range \\ 
 \hline\hline
 Normal & Hermite  & $(-\infty, \infty)$ \\
 \hline
 Uniform & Legendre  & $[-1, 1]$  \\
 \hline
 Exponential & Laguerre  & $[0, \infty)$ \\
 \hline
 Beta & Jacobi  & $(-1, 1)$ \\
 \hline
 Gamma & Generalized Laguerre  & $[0, \infty)$ \\
 \hline
\end{tabular}
\label{tab: orthog poly} 
\end{table}

To solve the PCE coefficients $\alpha_i$ in \eqref{Eqn: pce estimate}, the non-intrusive polynomial chaos (NIPC) method applies either the integration or regression approach. 
The integration approach makes use of the orthogonality property in \eqref{Eqn: orthogonal property} and projects the QoI onto each basis function, which results in a multi-dimensional integration problem:
\begin{equation}
\begin{aligned}
\label{eqn: integration}
    \alpha_i & = \frac{\left< f(U), \Phi_i \right> }{\left< \Phi_i^2 \right>} \\
    & = \frac{1}{\left< \Phi_i^2 \right>} \int_{\Gamma} f(u) \Phi_i(u) \rho(u) du\\
    & = \frac{1}{\left< \Phi_i^2 \right>}  \int_{\Gamma_1}\ldots \int_{\Gamma_d}  f(u_1, \ldots, u_d) \Phi_i(u_1,\ldots, u_d)\rho(u_1,\ldots, u_d)du_1 \ldots du_d. \\
\end{aligned}
\end{equation}
In practical problems, this integration problem is often approximated using numerical quadrature rules.


\subsection{Numerical quadrature rules}
This paper focuses on employing the integration-based NIPC method for solving multi-dimensional UQ problems.
In this context, the integration-based NIPC method is essentially solving the integration problem as in \eqref{eqn: integration}. 
We denote the integration problem as
\begin{equation*}
    I(f) = \int_{\Gamma}f(u) \rho(u)du,
\end{equation*}
where $u = (u_1, \ldots, u_d) \in \mathbb{R}^d$.
Solving this integration problem often involves utilizing the numerical quadrature approach, which approximates the integral as a weighted sum of function evaluations at specific quadrature points:
\begin{equation}
    I(f) \approx \sum_{i = 1}^n w^{(i)}f(u^{(i)}),
\end{equation}
where $u^{(i)} \in \Gamma$ and  $w^{(i)}> 0$ represent the nodes and weights, respectively, to be determined by the quadrature rule.

The objective of the quadrature rule is to effectively approximate a large set of functions with a minimal number of function evaluations. This is typically achieved by ensuring equality conditions hold for all functions $f$ within a polynomial subspace $\Pi$, as expressed by:
\begin{equation}
\label{eqn: exact integration}
    \int_{\Gamma}f(u) \rho(u)du = \sum_{i = 1}^n w^{(i)}f(u^{(i)}) \quad \text{for all} \quad  f \in \Pi_r.
\end{equation}
Here, we consider the cross-polynomial subspace with a total polynomial order of $r$, defined as:
\begin{equation}
    \Pi_r  = \text{span}\{ u^\alpha \ | \ \alpha \in \Lambda_r \},
\end{equation}
where
\begin{equation}
   \alpha = (\alpha_1, \ldots, \alpha_d),\quad  u^\alpha = \prod_{j = 1}^d (u_j)^{\alpha_j} \quad 
\end{equation}
and
\begin{equation}
    \Lambda_r = \left\{ \alpha  \mid  \sum_{i = 1}^d \alpha_i \leq  r \right\}.
\end{equation}
In many numerous integration problems, the evaluation functions exhibit smooth behavior and can be accurately approximated by the polynomials within the cross-polynomial subspace. In such instances, the quadrature rule enforcing Equation \eqref{eqn: exact integration} can prove to be highly effective.

\subsubsection{Gauss quadrature rule}
One approach to generating the quadrature points is through the use of the Gauss quadrature rule~\cite{golub1969calculation}.
In the one-dimensional case, the quadrature points $u^{(1)}, \ldots, u^{(k)}$ are the roots of the orthogonal polynomial $p_k(u)$ with degree $k$.
The polynomials here are defined in the same way as the PCE basis functions, and they also satisfy orthogonality:
\begin{equation}
    \left< p_i, p_j  \right> = \int_{\Gamma}  p_i(u) p_j(u)\rho(u) du = \delta_{ij}.
\end{equation}
The weights of the Gauss quadrature rule, $w^{(1)},\ldots, w^{(k)}$ can be computed by solving the moment-matching equations:
\begin{equation}
    \sum_{i=1}^{k} p_{j}\left(u^{(i)}\right) w^{(i)}=
    \left\{
    \begin{array}{ll}
        1 & \text { if } j=0 \\
        0 & \text { for } j=1, \ldots, k-1.
    \end{array}
    \right.
\end{equation}
In this approach, the univariate Gauss quadrature rule with $k$ nodes and weights can exactly integrate the univariate polynomials up to degree $2k -1$.

For multi-dimensional integrations, the Gauss quadrature rule extends the univariate rule to form a tensor structure, Assuming $k$ quadrature points in each dimension, the integral is approximated as 
\begin{equation}
   I \approx  \sum_{i_1 = 0}^{k} \ldots \sum_{i_d = 0}^{k} w_1^{(i_1)}\ldots w_d^{(i_d)} f(u_1^{(i_1)}, \ldots ,u_d^{(i_d)}) \rho(u_1^{(i_1)}, \ldots, u_n^{(i_d)}), 
\end{equation}
where $(u_i^{(1)},\ldots, u_i^{(k)})$ and $(w_i^{(1)},\ldots, w_i^{(k)})$ are the nodes and weights for the 1D Gauss quadrature rule in the $u_i$ dimension, respectively.
We represent the full-grid quadrature points as
\begin{equation}
\label{eqn: full tensor}
 \boldsymbol{u} = \boldsymbol{u}^{k}_1 \times \cdots \times \boldsymbol{u}^{k}_d,
\end{equation}
where $\boldsymbol{u}^{k}_i:= \{u_i^{(j)}\}_{j = 1}^k$, and $ \times $ represents the combination of quadrature points across dimensions using a tensor product.
Indeed, while this method allows exact integration of cross-polynomials up to a total order of $2k - 1$, it suffers from the curse of dimensionality. This is because the total number of quadrature points increases exponentially with the number of dimensions, as $n = k^d$.

To mitigate the curse of dimensionality the Gauss quadrature rule suffers from, the Smolyak sparse grid method~\cite{smolyak1963quadrature} offers a solution. 
This method strategically drops higher-order cross terms in the full-grid Gauss quadrature points, reducing the number of required quadrature points while preserving the accuracy of the quadrature rule to a significant extent. 
The sparse grid quadrature points can be expressed as:
\begin{equation}
\label{eqn：sparse grid}
    \boldsymbol{u} = \bigcup_{\ell-d+1 \leq|\mathbf{i}| \leq \ell}\left(\boldsymbol{u}^{i_{1}}_1 \times \cdots \times \boldsymbol{u}^{i_{n}}_d\right),
\end{equation}
where $\ell$ is the level of construction.
\subsubsection{Designed quadrature method}
For high-dimensional integration problems, a more effective quadrature rule is offered by the designed quadrature method proposed by Keshavarzzadeh et al.~\cite{keshavarzzadeh2018numerical}.
The core idea of the designed quadrature method is to numerically generate a quadrature rule that satisfies the multi-variate moment-matching equations:
\begin{equation}
\label{eqn: multi_mm}
    \sum_{i=1}^{n} \Phi_{\boldsymbol{i^\prime}} \left(u^{(i)}\right) w^{(i)}=\left\{\begin{array}{ll}
1 & \text { if } |\boldsymbol{i^\prime}|=0 \\
0 & \text { for } \alpha \in 0<|\boldsymbol{i^\prime}|<k,
\end{array}\right.
\end{equation}
where
\begin{equation}
    \pi_\alpha = \prod_{i=1}^d p_{\alpha_i}^{(i)}(u_i)
\end{equation}
and $p^{(i)}_j$ is the $j$-th orthogonal polynomial in the $u_i$ dimension.
If the nodes and weights of the quadrature rule satisfy the equations in \eqref{eqn: multi_mm}, then the quadrature rules can exactly integrate cross-term polynomials up to $(2k-1)$th order.
We denote the quadrature points and the weights as $\boldsymbol{u}: = (u^{(1)},\ldots, u^{(n)})$ and the weights $\boldsymbol{w}: = (w^{(1)},\ldots, w^{(n)})$ respectively. Then, the moment-matching equations in \eqref{eqn: multi_mm} can be written as residual equations:
\begin{equation}
    \mathcal{R} (\boldsymbol{u},\boldsymbol{w}) = 0, 
\end{equation}
where $\mathcal{R}:\mathbb{R}^{d \times n} \times \mathbb{R}^{n}  \to \mathbb{R}^n$.

The designed quadrature method approximates the solution of the moment-matching equations by solving an optimization problem formulated as
\begin{equation}
    \begin{aligned}
\min _{\boldsymbol{u}, \boldsymbol{w}} &\|\mathcal{R} (\boldsymbol{u},\boldsymbol{w})\|_{2} \\
\text { subject to } & u^{(j)} \in \Gamma, \quad j=1, \ldots, n, \\
& w^{(j)}> 0 , \quad j=1, \ldots, n.
\end{aligned}
\label{eqn: dq opt}
\end{equation}
The objective of this optimization, as shown in Equation \eqref{eqn: dq opt}, is to minimize the $L^2$ norm of the residual equations while constraining the nodes to lie within the support range and ensuring that the weights are positive.
This method has demonstrated superior effectiveness compared to the sparse-grid Gauss quadrature rule for many high-dimensional integration problems. The number of quadrature points required is typically chosen as $n = \eta\frac{(2d)^{k-1}}{(k-1)!}$, where $\eta \in [0.5,0.9]$ serves as a tuning parameter. The theoretical derivations for selecting the number of quadrature points, along with the strategies for optimization initialization, are thoroughly discussed in~\cite{keshavarzzadeh2018numerical}.

\subsubsection{Other popular numerical quadrature methods}
High-dimensional integration problems are prevalent in numerous fields beyond UQ. Over the years, many popular numerical quadrature methods have been developed to address these problems in various forms. In isogeometric analysis, B-spline quadrature rules have been found to be more efficient than Gauss quadrature rules~\cite{johannessen2017optimal, bartovn2017gauss, calabro2017fast, bartovn2020efficient}, as they can approximate smooth functions using fewer basis functions. However, their application with PCE  for solving UQ problems has only recently been explored~\cite{rahman2020spline}. Additionally, quasi-Monte Carlo methods~\cite{niederreiter1992random, sloan1998quasi} represent another class of widely used numerical quadrature methods. Unlike Gauss quadrature, which assesses accuracy based on polynomial exactness of integration, quasi-Monte Carlo methods evaluate integration accuracy based on the discrepancy properties of the sample points.

\subsection{Computational graph transformations for efficient tensor-grid evaluations}

\begin{figure}%
    \centering
    \subfloat[\centering Computational graph without using AMTC]{{\includegraphics[width=5cm]{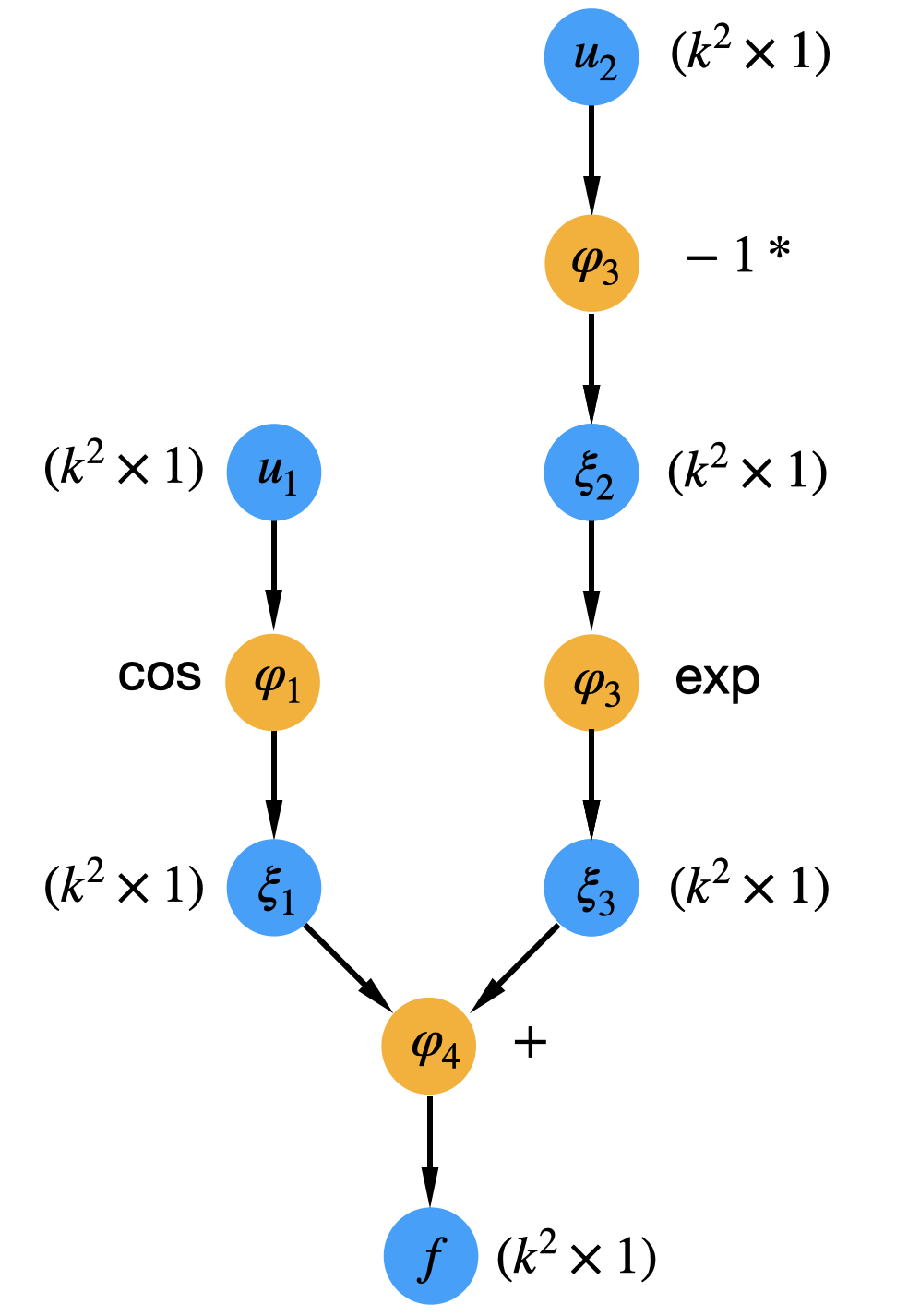} }}%
    \qquad
    \subfloat[\centering Computational graph using AMTC]{{\includegraphics[width=5cm]{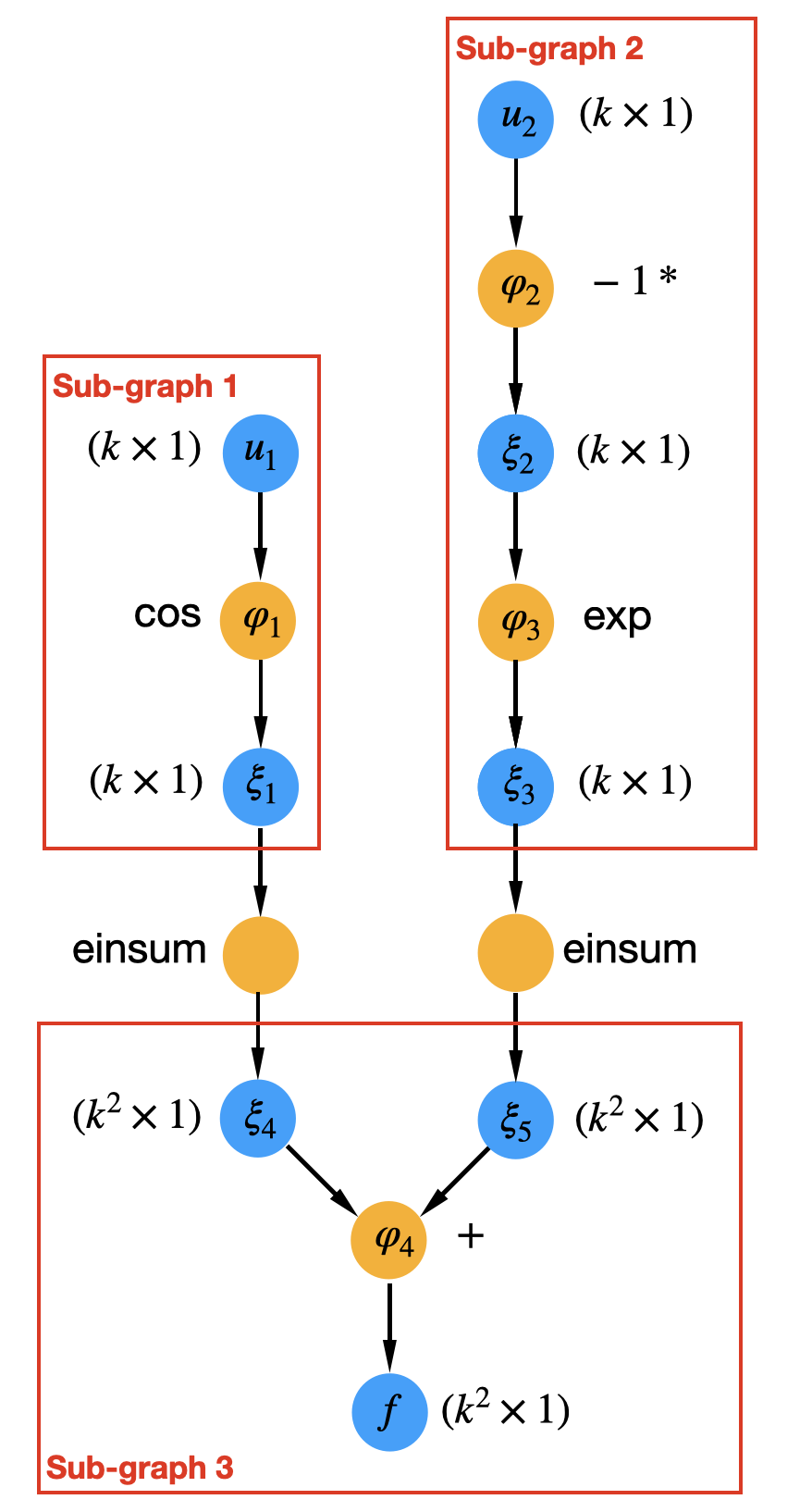} }}%
    \caption{Computational graphs with data size for full-grid input points evaluation on $f = cos(u_1) + exp(-u_2)$~\cite{wang2023accelerating}}%
    \label{fig:graph comparison}%
\end{figure}
\textit{Accelerated Model evaluations on Tensor grids using Computational graph transformations} (AMTC) is a computational graph transformation method, recently introduced by Wang et al. in \cite{wang2023accelerating}, 
with its core idea initially proposed in \cite{wang2022efficient}.
AMTC accelerates the evaluation time of a computational model on tensor-grid inputs by eliminating repeated evaluations on the operation level. 
Any computational model can be represented as a computational graph with basic operations. 
In a conventional for-loop approach, when evaluating a computational model on tensor-grid inputs, each operation on the computational graph needs to be evaluated at every input point. 
However, with tensor-grid inputs, where inputs possess a fully tensorial structure, only a small set of unique values exists in each input space. 
Consequently, many operations within the computational model might depend on only a few inputs, leading to many redundant repeated evaluations at the operation level with a for-loop approach.
To address this inefficiency, AMTC eliminates the redundant evaluations by generating a modified computational graph.
The modified computational graph is partitioned into smaller sub-graphs based upon on which inputs each operation depends.
Operations within each sub-graph share the same input space and are solely evaluated on distinct nodes within their input space.
To ensure proper data flow between these sub-graphs, Einstein summation (\textit{Einsum}) operations are inserted at each connection joint between sub-graphs. 

To consider an illustrative example,
assume we want to evaluate a simple function,
\begin{equation}
f = \cos(u_1) + \exp(-u_2),
\end{equation}
at full-grid input points with $k$ input points at each dimension.
The given function can be represented as a sequence of elementary operations:
\begin{equation}
\begin{aligned}
& \xi_1 = \varphi_1 (u_1) = \cos (u_1); \\
& \xi_2 = \varphi_2 (u_2) = - u_2; \\
& \xi_3 = \varphi_3 (\xi_2) = \exp (\xi_2); \\
& f = \varphi_4 (\xi_1, \xi_3) = \xi_1 + \xi_3. \\
\end{aligned}
\end{equation}
The tensor-grid input points are denoted as
\begin{equation}
\boldsymbol{u} = \begin{Bmatrix}
(u_1^{(1)}, u_2^{(1)}) & \ldots & (u_1^{(k)}, u_2^{(1)}) \\
\vdots & \ddots & \vdots \\
(u_1^{(1)}, u_2^{(k)}) & \ldots & (u_1^{(k)}, u_2^{(k)}) \\
\end{Bmatrix}.
\end{equation} 
The computational graphs to perform these full-grid evaluations with and without the AMTC method are shown in Fig.~\ref{fig:graph comparison}.
In comparison, without AMTC, each operation in the computational graph is evaluated $k^2$ times, corresponding to the total number of input points. However, by applying AMTC, operations that depend only on $u_1$ and $u_2$ are only evaluated $k$ times since their outputs have only $k$ distinct values. 

AMTC has been integrated into the middle end of the compiler for the \textit{Computational System Design Language} (CSDL)~\cite{gandarillas2022novel}, a recently developed domain-specific language tailored for tackling multidisciplinary design, analysis, and optimization problems. 
CSDL has gained significant popularity in addressing large-scale multidisciplinary design optimizations, with applications including designing novel air-taxi~\cite{sarojini2023large, ruh2024large} and spacecraft~\cite{gandarillas2023talos} concepts. 
Within the CSDL compiler, the AMTC serves as a middle-end algorithm, automatically generating the modified computational graph to minimize the model evaluation cost on tensor-grid inputs.

This method has been used in combination with the integration-based NIPC method, which forms full-grid quadrature rules for low-dimensional UQ problems. We refer to the combination of integration-based NIPC and AMTC as \textit{graph-accelerated NIPC}, and this method has demonstrated exceptional effectiveness in solving certain low-dimensional UQ problems involving multidisciplinary or multi-point systems~\cite{wang2024graphthesis}.

\section{Methodology}
\label{Sec: Methodology}
This paper proposes a new framework that generates a tensor-structured quadrature rule that is specifically tailored for a given computational model. This quadrature rule is intended to be used with the graph-accelerated NIPC method to efficiently solve the UQ problem.
\subsection{Generation of tensor-structured quadrature rules}
We use the degree of polynomials that a quadrature rule can exactly integrate as the measure of its accuracy. Under this criterion, if we maintain a non-tensorial structure for the quadrature rule,  the most effective strategy is to utilize the designed quadrature method. In this scenario, the quadrature points can be expressed as:
\begin{equation}
    \boldsymbol{u} = \boldsymbol{u}^{n}_{\{1,\ldots, d\}}.
\end{equation}
The nodes and weights in the quadrature rule are determined by solving the optimization problem outlined in \eqref{eqn: dq opt}, and the number of quadrature points is chosen as
\begin{equation}
\label{eqn: dp num}
    n = \eta\frac{(2d)^{k-1}}{(k-1)!}, \quad \eta \in [0.5,0.9].
\end{equation}
Adopting this method ensures that the resulting quadrature rule can effectively integrate polynomials up to $(2k-1)$th order.
If a fully tensorial structure for the quadrature points is necessary, the optimal method to employ is the full-grid Gauss quadrature method.
In this approach, the quadrature points are expressed as:
\begin{equation}
 \boldsymbol{u} = \boldsymbol{u}^{k}_{\{1\}} \times \cdots \times \boldsymbol{u}^{k}_{\{d\}},
\end{equation}
where $\boldsymbol{u}^{k}_{\{i\}} $ represents the 1D Gauss quadrature points in $u_i$ dimension. 
The sets of 1D Gauss quadrature points in each dimension are determined separately to ensure the nodes and weights can exactly integrate the univariate polynomials up to $(2k-1)$th order. 
Maintaining a fully tensorial structure of the 1D Gauss quadrature points and weights ensures that the full-grid quadrature rule can exactly integrate cross-polynomials up to the $(2k-1)$th order, albeit while requiring $n = k^d$ quadrature points.
In fact, in each dimension, the 1D Gauss quadrature points and weights correspond to the global minimum of the optimization problem outlined in \eqref{eqn: dq opt}. 
Assuming that the optimization problem in \eqref{eqn: dq opt} always finds the global minimum, employing the full-grid Gauss quadrature rule is equivalent to applying the designed quadrature method in each dimension and subsequently forming a fully tensorial structure between the 1D solutions.

In addition to the fully tensorial and non-tensorial structure options, we can also preserve a partially tensorial structure for the quadrature points while ensuring accurate integration of all polynomials up to a specified order. For instance, in a 4D integration problem, we could maintain a partially tensorial structure as:
\begin{equation}
 \boldsymbol{u} = \boldsymbol{u}^{n_1}_{\{1,2\}} \times  \boldsymbol{u}^{n_2}_{\{3,4\}},
\end{equation}
where a tensorial structure is retained between the quadrature points in the space of $u_1, u_2$ and the quadrature points in the space of $u_3, u_4$.
If the quadrature rules in each space are selected such that the quadrature points can precisely integrate polynomials up to the same order, then this partially tensorial quadrature rule can also precisely integrate the polynomials to the same order in the space of $u$.
The total number of tensorial structure options for the quadrature rule can be represented by the Bell number from combinatorial mathematics, which is given by:
\begin{equation}
\label{eqn: bell_num}
    B_{d}=\sum_{i=0}^{d}\left\{\begin{array}{l}
i \\
d
\end{array}\right\}.
\end{equation}
For example, for $d = 3$, $B_{3} = 5$, the five options of the tensorial structure are:
\begin{equation}
\begin{aligned}
\boldsymbol{u} &  = \boldsymbol{u}^{n}_{\{1,2,3\}}, \\
\boldsymbol{u} &  = \boldsymbol{u}^{n_1}_{\{1,2\}} \times  \boldsymbol{u}^{n_2}_{\{3\}}, \\
\boldsymbol{u} &  = \boldsymbol{u}^{n_1}_{\{1,3\}} \times  \boldsymbol{u}^{n_2}_{\{2\}}, \\
\boldsymbol{u} &  = \boldsymbol{u}^{n_1}_{\{2,3\}} \times  \boldsymbol{u}^{n_2}_{\{1\}}, \\
\boldsymbol{u} &  = \boldsymbol{u}^{n_1}_{\{1\}} \times  \boldsymbol{u}^{n_2}_{\{2\}} \times  \boldsymbol{u}^{n_3}_{\{3\}}.
\end{aligned}    
\end{equation}
\begin{figure}%
    \centering
    \subfloat[\centering $\boldsymbol{u}  = \boldsymbol{u}^{n}_{\{1\}} \times  \boldsymbol{u}^{n}_{\{2\}} \times  \boldsymbol{u}^{n}_{\{3\}}$]{{\includegraphics[width=4.5cm]{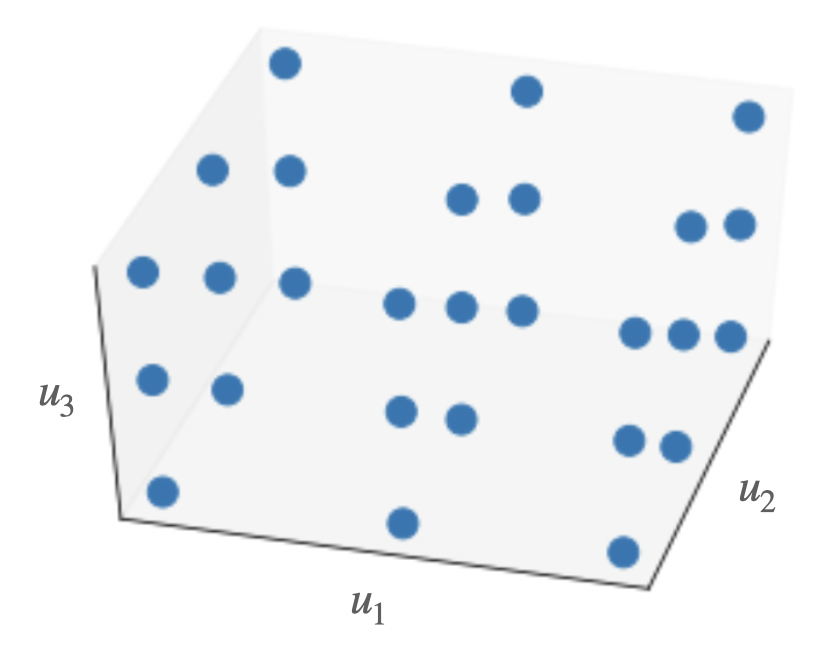} }}%
    \qquad
    \subfloat[\centering $\boldsymbol{u}  = \boldsymbol{u}^{n_1}_{\{1,2\}} \times  \boldsymbol{u}^{n_2}_{\{3\}}$]{{\includegraphics[width=4.6cm]{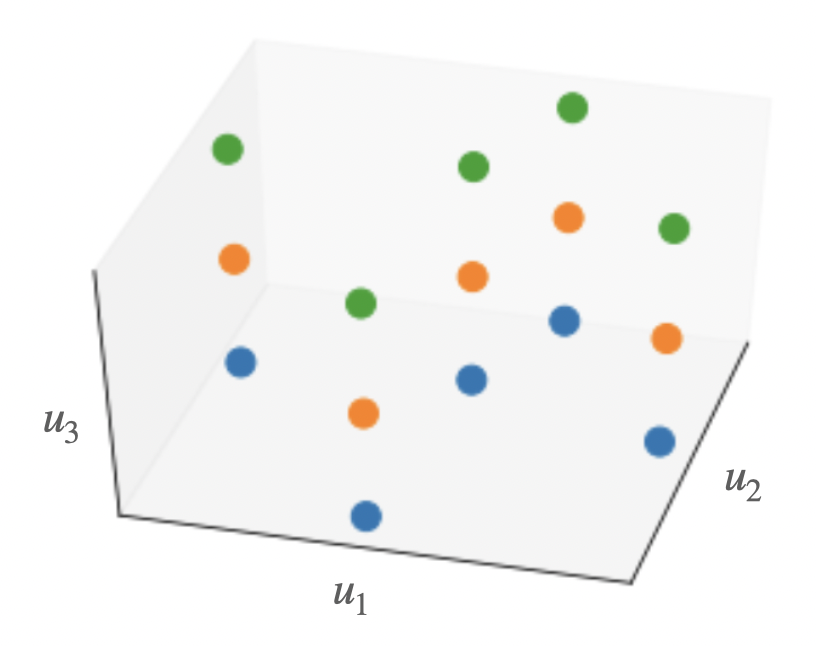} }}%
     \qquad
    \subfloat[\centering $\boldsymbol{u}   = \boldsymbol{u}^{n}_{\{1,2,3\}}$]{{\includegraphics[width=4.8cm]{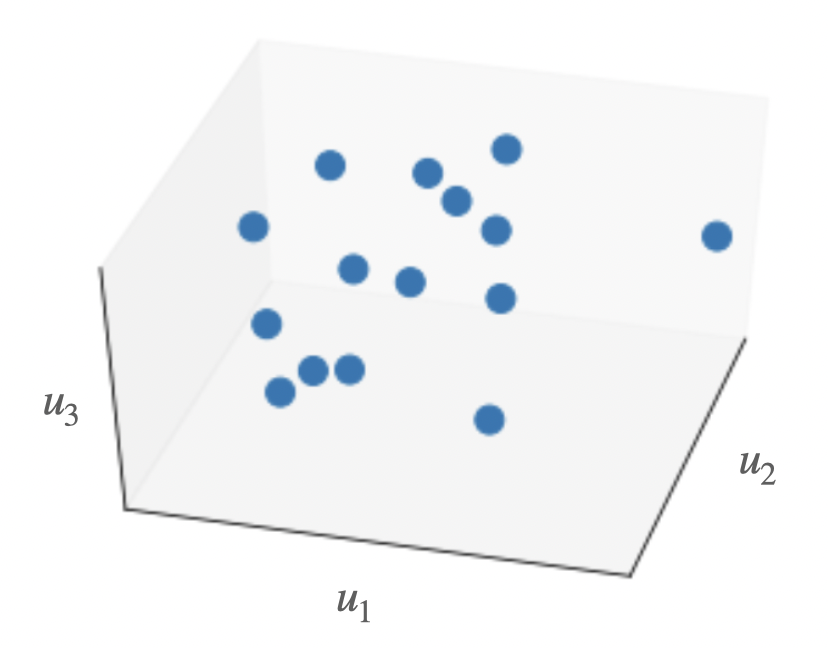} }}%
    \caption{Visualizations of quadrature points with different tensorial structure options}%
    \label{fig:qp visual comparison}%
\end{figure}
Visualizations of quadrature points with different tensorial structures can be found in Fig.~\ref{fig:qp visual comparison}. 

Given a specific tensorial structure option for the quadrature rule, we need to decide on the method for generating the quadrature points within each space and determine the requisite number of quadrature points to attain the desired level of accuracy.
In our method, for a particular tensorial structure of the quadrature rule, we employ the designed quadrature method in each space. This method determines the nodes and weights to accurately integrate polynomials in that space up to a specified order.
Adapted from the Gauss quadrature rule, we use the level of accuracy $k$ for a quadrature rule to indicate its capability of precisely integrating polynomials up to the  $(2k-1)$th order in the integration space.
The number of quadrature points utilized in the designed quadrature method is given by \eqref{eqn: dp num}, tailored for high-dimensional integration problems.
In our setting, we also utilize the designed quadrature method for relatively low-dimensional spaces. Therefore, we choose the required number of quadrature points using:
\begin{equation}
\label{eqn: opt. num}
    n = \left\{\begin{array}{ll}
\frac{k^d}{d} & \text { for } d \leq 2 \\
0.9\frac{(2d)^{k-1}}{(k-1)!} & \text { for } d > 2.
\end{array}\right.
\end{equation}
When the number of dimensions is greater than two, we follow the designed quadrature rule in \eqref{eqn: dp num} with $\eta = 0.9$.
However, for 1D and 2D cases, we choose $n =\frac{k^d}{d} $. In this way, in the 1D space, it chooses $k$ quadrature points to achieve $k$ level of accuracy, which matches the Gauss quadrature rule.
Consequently, for the  1D space, we can avoid solving the optimization problem in the designed quadrature method and directly use the 1D Gauss quadrature rule, as it represents the exact solution of the designed quadrature method.
In 2D space, we have $n = \frac{k^2}{2}$, this has been numerically tested effective to generate sufficient quadrature points to achieve $k$ level of accuracy for the quadrature rule.
This approach results in generating the full-grid Gauss quadrature rule if a fully tensorial structure option is chosen, and generating the designed quadrature rule if a non-tensorial structure option is chosen.

\subsection{Tensor-grid evaluations with AMTC}
When not using the AMTC method for model evaluations, tensor-structured quadrature rules often underperform compared to non-tensorial quadrature rules. This is because tensor-structured quadrature rules always require evaluations of the model at more input points than non-tensorial quadrature rules to achieve the same level of accuracy. 
However, when utilizing the computational graph transformation method, AMTC, redundant operations incurred by the tensorial structure of the input points are eliminated at the operation level. 
Consequently, the overall cost of evaluating the model is no longer directly tied to the number of quadrature points utilized.
This gives us the reason for devising a custom tensor-structured quadrature rule tailored for a given computational model.

\begin{figure}
    \centering
    \subfloat[\centering $\boldsymbol{u}  = \boldsymbol{u}^{k}_{\{1\}} \times  \boldsymbol{u}^{k}_{\{2\}} \times  \boldsymbol{u}^{k}_{\{3\}}$]{{\includegraphics[width=6cm]{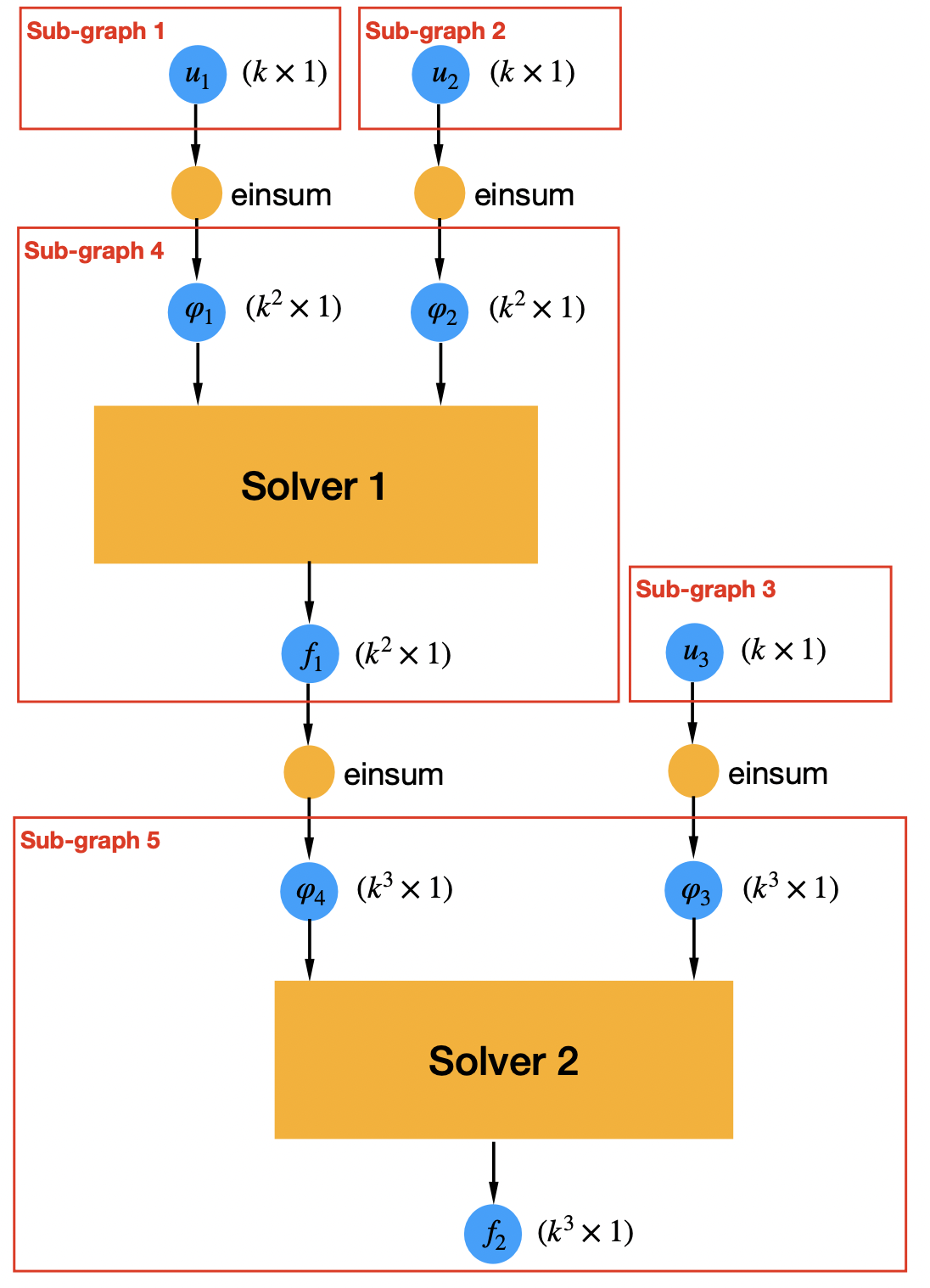} }}%
    \qquad
    \subfloat[\centering $\boldsymbol{u}  = \boldsymbol{u}^{k_1}_{\{1,2\}} \times  \boldsymbol{u}^{k_2}_{\{3\}}$]{{\includegraphics[width=6cm]{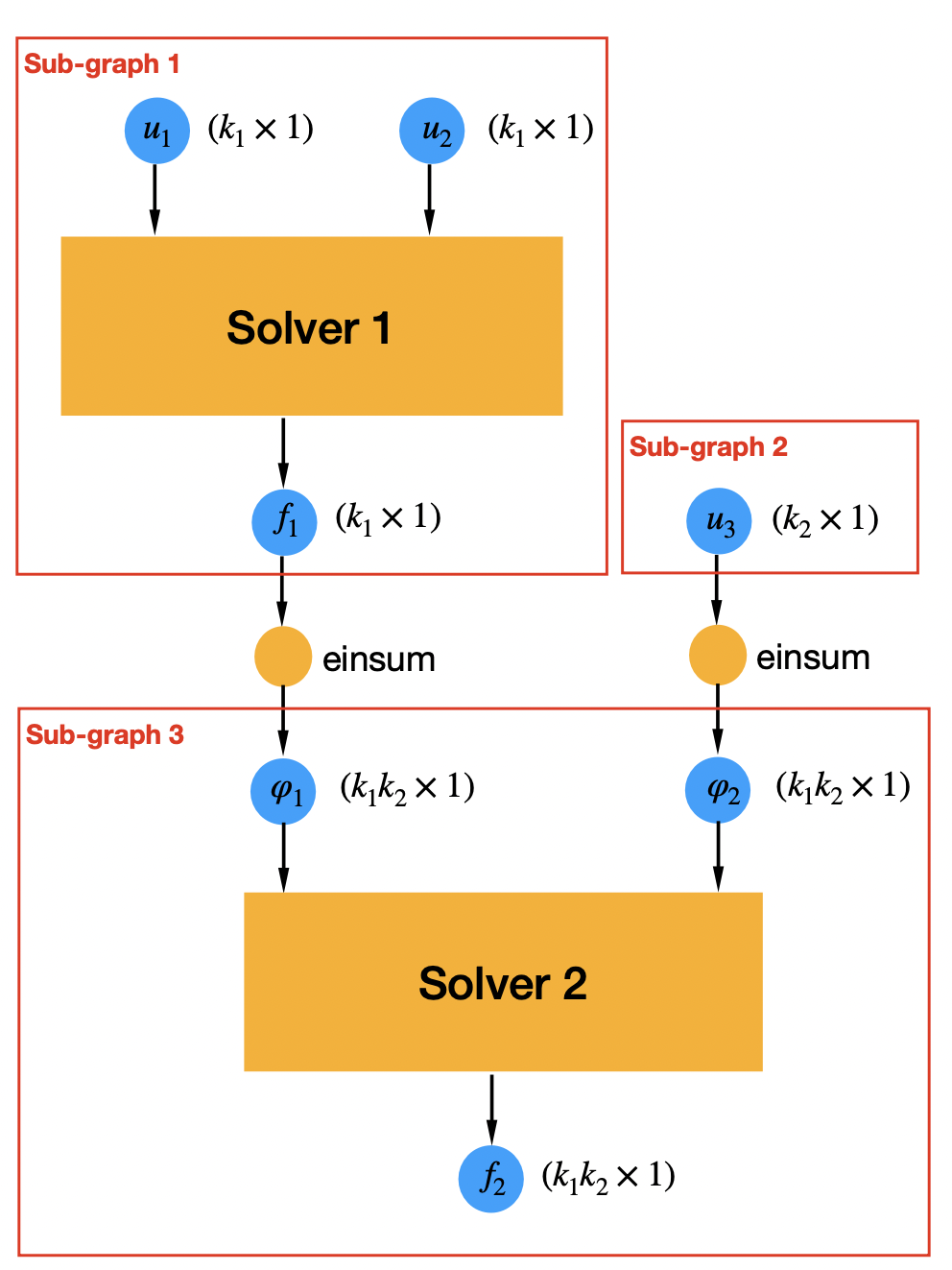} }}%
    \caption{Computational graphs after AMTC on fully and partially tensor structured input points}%
    \label{fig: partial graph comparison}%
\end{figure}

For instance, consider a 3D UQ problem involving two computationally expensive solvers. The computational model can be written as
\begin{equation}
\label{eqn: example_solvers}
\begin{aligned}
& f_1 = \mathcal{F}_1 (u_1, u_2), \\
& f_2 = \mathcal{F}_2 (f_1, u_3),\\
\end{aligned}
\end{equation}
where $\mathcal{F}_1$ and $\mathcal{F}_2$ represent the first and second solver, respectively, and the UQ problem aims to compute the risk measures of $f_2$ under the effect of the three uncertain inputs. 
In this case, we have two uncertain inputs associated with solver 1, and one uncertain input associated with solver 2.
Consequently, the operations in solver 1 are dependent on $u_1$ and $u_2$ while the operations in solver 2 are dependent on $u_1$, $u_2$, and $u_3$.
In such scenario, an optimal quadrature rule can be easily formulated following a partially tensorial structure:
\begin{equation}
    \boldsymbol{u}  = \boldsymbol{u}^{k_1}_{\{1,2\}} \times \boldsymbol{u}^{k_2}_{\{3\}}.
\end{equation}
The corresponding model evaluations can be significantly accelerated using the AMTC method. To demonstrate, we show the computational graphs when performing model evaluations on fully tensorial quadrature points $\boldsymbol{u}^{k}_{\{1\}} \times \boldsymbol{u}^{k}_{\{2\}} \times \boldsymbol{u}^{k}_{\{3\}}$ and partially tensorial quadrature points $\boldsymbol{u}^{k_1}_{\{1,2\}} \times \boldsymbol{u}^{k_2}_{\{3\}}$ after applying AMTC in Fig.~\ref{fig: partial graph comparison}.
We observe that, in both cases, AMTC generates modified computational graphs that reduce the number of evaluations on solver 1. In comparison, the partially tensor-structured quadrature rule is more favorable than the fully tensor-structured quadrature rule, as both the $u_1$ and $u_2$ affect the same operations on the computational graph, so there is no need to maintain a tensor structure between $u_1$ and $u_2$ like in the fully tensor-structured quadrature rule. For the partially tensor-structured quadrature rule, by grouping the space of $u_1$ and $u_2$, the total number of quadrature points in this space, $k_1$, is smaller than the full-grid quadrature rule which has $k^2$ quadrature points in this space. Consequently, utilizing the partially tensor-structured quadrature rule results in fewer evaluations for both solver 1 and solver 2. 

\subsection{Finding a desirable tensor structure option}

For specific UQ problems, particularly those involving multidisciplinary systems, certain partially tensor-structured quadrature rules demonstrate superior performance compared to existing ones when utilized with the graph-accelerated NIPC method. However, identifying an optimal tensor-structure option may be challenging.

Theoretically, we can estimate the model evaluation costs after applying AMTC for each tensor-structured rule.
To do that, for a computational model $f = \mathcal{F}(u)$, we denote the data and operations nodes as $\xi_i$ and $\varphi_j$, respectively. Here $l$ represents the number of operations in the computational graph. For example, consider the function
\begin{equation}
\label{eqn: simple}
    f = cos(u_1) + exp(-u_2),
\end{equation}
with its corresponding computational graph described as
\begin{equation}
\label{eqn: computational process}
    \begin{aligned}
    & \xi_1 =  \varphi_1 (u_1) = cos (u_1); \\
    & \xi_2 = \varphi_2 (u_2) =  - u_2; \\
    & \xi_3 =  \varphi_3 (\xi_1) =  exp (\xi_2); \\
    & f = \varphi_4 (\xi_2, \xi_3) =  \xi_2 + \xi_3. \\
    \end{aligned}
\end{equation}
From the computational graph, we can generate the dependency information which stores whether the output of one operation is dependent on one input as a binary number, written as $D(\varphi_i, u_j)$. The dependency information can be viewed in a matrix. 
The dependency matrix for the function \eqref{eqn: simple} is presented in Tab. \ref{tab: dep inf}, which indicates the dependencies between operations and input variables.
\begin{table}[]
\caption{Operations dependency information for function $f = cos(u_1) + exp(-u_2)$}
\centering
\begin{tabular}{c | c c} 
 $D(\varphi_i, u_j)$ & $u_1$ & $u_2$ \\ 
 \hline
 $\varphi_1$ & 1 & 0 \\

 $\varphi_2$ & 0  & 1  \\

 $\varphi_3$ & 0  & 1 \\

 $\varphi_4$ & 1  & 1 \\
\end{tabular}
\label{tab: dep inf} 
\end{table}
To estimate the evaluation cost of each tensorial structure for the quadrature rule to achieve the $k$-level of accuracy, we also need to store the information regarding the evaluation cost of each operation on one quadrature point. 
This information can be represented either by the total number of floating-point operations (FLOPs) or the evaluation time.
For the non-tensorial structure, if we assume all operations are dependent on at least one input, no repeated evaluations can be reduced by the AMTC method. 
Assume that the operations' cost on all of the quadrature points is roughly the same, the total model evaluation cost can be approximated as:
\begin{equation}
    \text{Cost} \approx n O\left(f(u)\right) \approx n \sum_{i=1}^l O(\varphi_i),
\end{equation}
where $O(\cdot)$ represents the evaluation cost of the function/operation and the number of quadrature points $n$ is chosen as in \eqref{eqn: opt. num}. 
If we maintain a full-grid structure with $k$ quadrature points in each dimension, the evaluation cost without the AMTC method is 
\begin{equation}
    \text{Cost} \approx k^d O\left(\mathcal{F}(u^{(1)})\right) \approx k^d \sum_{i=1}^l O(\varphi_i).
\end{equation}
However, with the AMTC method, each operation in the graph is evaluated only for distinct values. In the full-grid case, if the output of an operation is dependent on $\tilde{d}$ number of uncertain inputs, there are only $k^{\tilde{d}}$ quadrature points in its dependent inputs space, and that operation will be evaluated for $k^{\tilde{d}}$ number of times using the AMTC method.
In this case, we omit the cost of the Einstein summation nodes added to the computational graph by the AMTC method, the evaluation cost can be estimated as
\begin{equation}
    \text{Cost}  \approx  \sum_{i=1}^l k^{\sum_{j=1}^d D(\varphi_i, u_j)} O(\varphi_i).
\end{equation}
Using the same idea, we can also estimate the model evaluations cost for any tensorial structure option of the quadrature points. 
For example, consider the setting where we have a partially tensorial structure for the quadrature points, such as
\begin{equation}
\boldsymbol{u} = 
\boldsymbol{u}^{k_1}_{\{1,2\}}
\times
\boldsymbol{u}^{k_2}_{\{3\}}.
\end{equation}
In our method, $k_1$ and $k_2$ are chosen from \eqref{eqn: opt. num} using $d = 1$ and $d = 2$ respectively.
In this case, there are $k_1$ quadrature points in the space of $u_1$ and $u_2$, and the $k_2$ quadrature points in the space of $u_3$. 
With the AMTC method, for the operations that are dependent on only $u_1$, only $u_2$ or 
only $u_1$ and $u_2$, they will be evaluated $n_1$ times. 
Operations dependent on only $u_3$, will be evaluated $n_2$ times. Operations not dependent on any input are evaluated once. The rest of the operations are evaluated $n_1n_2$ times. The model evaluations cost for this tensor-structured quadrature rule can be estimated as
\begin{equation}
    \text{Cost}  \approx  \sum_{i=1}^l k_1^{D(\varphi_i, u_1)D(\varphi_i, u_2)} k_2^{D(\varphi_i, u_3)} O(\varphi_i).
\end{equation}

Estimating the total model evaluation cost for all tensor structure options enables us to choose the optimal one with the lowest estimated cost. However, the challenge lies in the fact that the total number of tensor structure options follows the Bell number as shown in \eqref{eqn: bell_num}, which scales significantly with the problem dimensions. To avoid this significant overhead cost associated with numerically finding the optimal tensor structure option, we propose an alternative approach. 
This approach involves selecting a desirable tensor structure option based on our understanding of the computational model and an analysis of the computational graph. 
By leveraging information from the computational graph and characteristics of the computational model, we can make informed decisions to choose a tensor structure that is likely to lead to efficient model evaluations. 
This approach balances computational efficiency with accuracy, making it a practical solution for many UQ problems with complicated computational models.

When dealing with computational models involving multiple disciplines or solvers, identifying a desirable tensor-structure option can sometimes be achieved by examining the computational graph at the solver level. For instance, in the example provided in \eqref{eqn: example_solvers}, we can easily deduce the optimal tensor structure based on insights gained from the computational graph.
In other cases, we can perform some analysis on the computational graph to help us decide on the desired tensor-structure option.
Our rationale here is based on the observation that complex computational models typically contain one or two computationally expensive operations, such as linear or non-linear solvers, which significantly contribute to the overall computational cost. 
We aim to identify uncertain inputs that these expensive operations do not depend on and then establish a tensor structure to minimize the number of evaluations required for these operations, with the utilization of AMTC.
To do that, we define the \textit{sparsity ratio} (SR) of an uncertain input as the ratio of its dependent operations' cost over the total model evaluation cost which can be approximated as
\begin{equation}
\label{eqn: sr}
    \text{SR}(u_i) \approx \frac{\sum_{i= 1}^lO(\varphi_i) D(\varphi_i, u_j)}{O(f(u))},
\end{equation}
where $\text{SR}(u_i)$ represents the sparsity ratio of $u_i$ and it measures the percentage of the model evaluation cost that is dependent on $u_i$.
We identify the sparse uncertain inputs based on the condition that their sparsity ratio is $< 5\%$.
We then partition the uncertain inputs as 
\begin{equation}
    u  = (u_{s}, u_{ns}), 
\end{equation}
where $u_{s}  = (u_{s_1}, \ldots, u_{s_{d_1}})\in \mathbb{R}^{d_1}$ is the set of sparse uncertain inputs and $u_{ns} = (u_{ns_1}, \ldots, u_{ns_{d_2}}) \in \mathbb{R}^{d_2}$ is the set of non-sparse uncertain inputs with $d = d_1 + d_2$. 

In this case, if all of the uncertain inputs are sparse uncertain inputs, we may want to form a full-grid quadrature rule following the Gauss quadrature method, such that the quadrature points can be written as
\begin{equation}
\boldsymbol{u} = 
\boldsymbol{u}^{k}_{\{1\}}
\times \ldots \times
\boldsymbol{u}^{k}_{\{d\}},
\end{equation}
where $\boldsymbol{u}^{k}_{\{i\}}$ represents the $k$ quadrature points in $u_{i}$ space.
In this way, even though the number of quadrature points increases exponentially as $n = k^d$, the model evaluation cost can be significantly accelerated using the AMTC method which only evaluates each operation on the distinct quadrature points in the space of the uncertain inputs on which it depends.
However, in most cases, for most UQ problems, there often only exist a few sparse uncertain inputs. In this case, we suggest choosing the quadrature rule that maintains a tensor structure between the quadrature points in the sparse uncertain input space and quadrature points in the space of non-sparse uncertain inputs, such that the quadrature points can be written as
\begin{equation}
\boldsymbol{u} = 
\boldsymbol{u}^{k_1}_{\{ns\}}
\times
\boldsymbol{u}^{k_2}_{\{s\}},
\end{equation}
where $\boldsymbol{u}^{k_1}_{\{ns\}}$ represents quadrature points in $u_{ns}$ space with $k_1$ points and $\boldsymbol{u}^{k_2}_{\{s\}}$ represent the quadrature points in $u_{{s}}$ space with $k_2$ points.
This results in a total of $k_1k_2$ quadrature points in the quadrature rule. 
However, when using the AMTC method to modify the computational graph, the computational cost on the modified computational graph would scale roughly linearly with $k_1$ as most of the expensive operations in the computational graph would only depend on $u_{ns}$ and would only be evaluated $n_1$ times.
An alternative option is to formulate the quadrature rule as
\begin{equation}
\label{eqn: qp_total}
\boldsymbol{u} = 
\boldsymbol{u}^{k_1}_{\{ns\}}
\times
\boldsymbol{u}^{k}_{\{s_1\}}\times \ldots,\boldsymbol{u}^{k}_{\{s_{d_1}\}},
\end{equation}
which forms a tensor structure between the space of each sparse uncertain input.
This option results in more quadrature points but it is easier to generate the quadrature rule and can be more effective to use if we know the sparse uncertain inputs are associated with different solvers.

\begin{figure}[hbt!]
\centering
  \includegraphics[width= 5 cm]{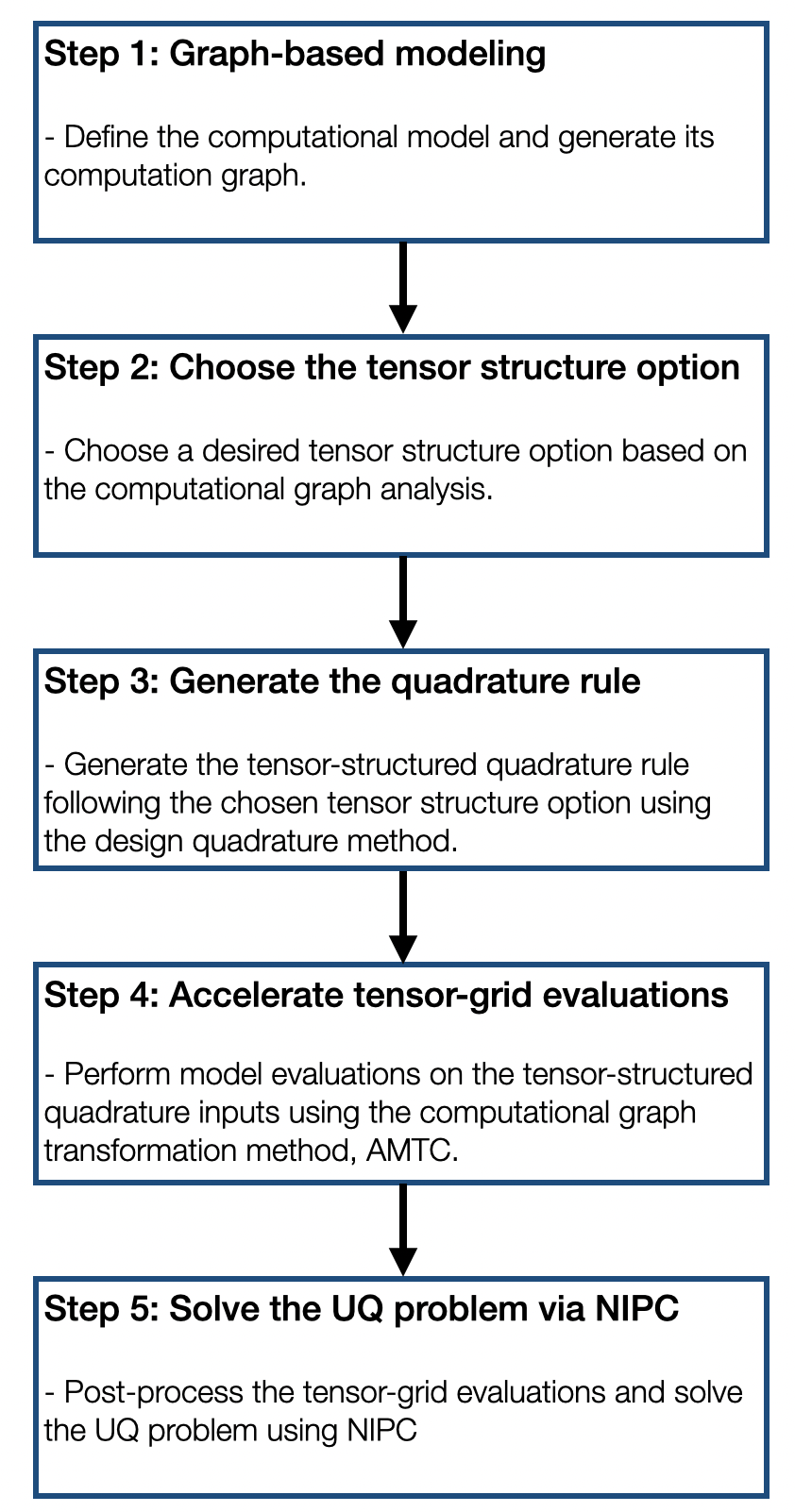}
\caption{A flow chart showing how the partially tensor-structured quadrature rule can be used with graph-accelerated NIPC to solve an UQ problem.}
\label{fig:flowchart}
\end{figure}

We detail the specific steps of applying this method with graph-accelerated NIPC in Algorithm~\ref{alg: partial tensor}, and a corresponding flowchart is provided in Fig.~\ref{fig:flowchart}.
\begin{algorithm}
        \setstretch{1.17}
        \small
        \caption{Partially tensor-structured quadrature rule for graph-accelerated NIPC}
        \label{alg: partial tensor}
        \begin{algorithmic}[1]
        \State Specify a computational graph for a numerical model and the uncertain inputs, $U$
        \State Specify the level of accuracy of the quadrature rule, $k$
        \State Choose a desired tensor structure option based on the analysis of the computational graph
        \State Compute the required number of quadrature points in each space following~\eqref{eqn: dp num} 
        \State Compute the quadrature points and weights in each space by solving~\eqref{eqn: dq opt}
        \State Formulate the quadrature rule using the chosen tensorial structure
        \State Perform the model evaluations on the tensor-structured quadrature points using AMTC 
        \State Post-process the evaluations using NIPC to solve the UQ problem 
        \end{algorithmic}
\end{algorithm}
\section{Numerical Results}
\label{Sec: Numerical Results}
\subsection{Laser-beam powered UAV multidisciplinary model}
\begin{figure}[hbt!]
\centering
  \includegraphics[width= 8 cm]{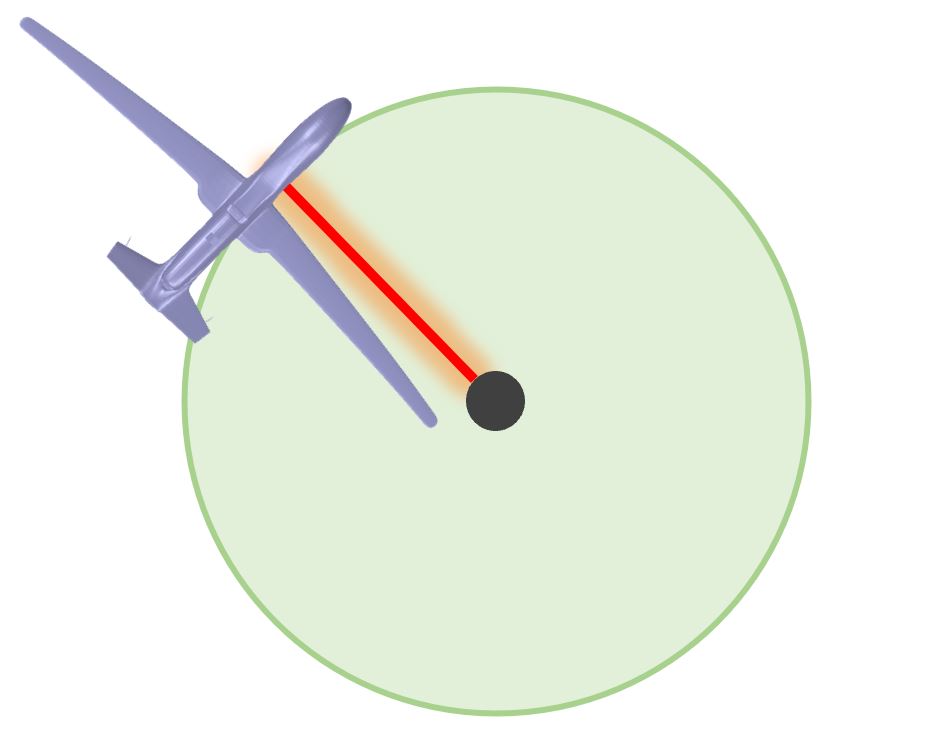}
\caption{A circular cruise mission around a ground station}
\label{fig:mission plot}
\end{figure}
\begin{table}[h!]
    \centering
    \begin{tabular}{c | c }
         Uncertain inputs & Distributions \\
         \hline
         \text { Payload weight $W_p$ ($kg$)} & {$\mathcal{N}(90, 10)$} \\
         \text {Atmospheric extinction coefficient $\sigma$} & {$\mathcal{N}(0.2,0.02)$} \\
         \text {Flight altitude $h$ (m)} & {$\mathcal{N}(10,000, 1000)$} \\
         \text {Flight velocity $v$ ($m/s$)} & {$\mathcal{N}(100, 10)$} \\
    \end{tabular}
    \caption{Uncertain inputs' distributions for the UAV problem}
    \label{tab:beam_uncertain_inputs}
\end{table}
\begin{figure}[hbt!]
\centering
  \includegraphics[width= 16 cm]{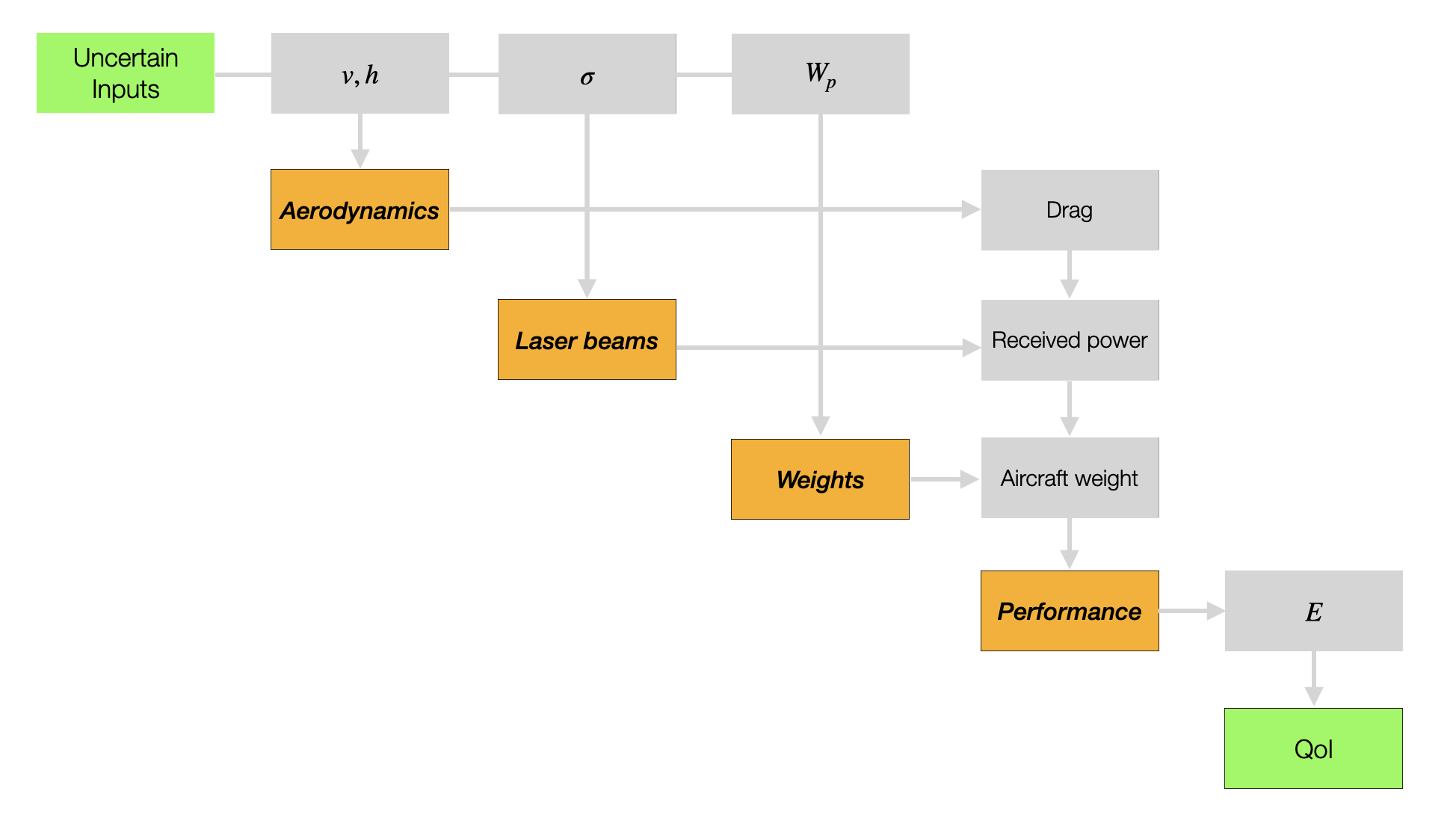}
\caption{Multidisciplinary structure of the UAV model}
\label{fig:multidisp diagram uav}
\end{figure}

The first test problem is a 4D UQ problem adapted from a UAV design scenario~\cite{orndorff2023gradient}. 
This computational model assesses the total energy stored during a cruise mission undertaken by a laser-beam-powered UAV as it orbits a ground station for one complete cycle while being recharged by the laser beam emitted from the ground station. 
The mission is demonstrated in Fig. \ref{fig:mission plot}.
We consider four uncertain inputs associated with the operating conditions, which are flight velocity, flight altitude, payload weight, and atmospheric extinction coefficient. 
Their distributions are shown in Tab.~\ref{tab:beam_uncertain_inputs}.
The objective of this UQ problem is to compute the expectation of the stored energy considering the influence of these uncertain inputs.

The computational model used in this problem involves four disciplines: aerodynamics, laser beams, weights, and performance. The multidisciplinary structure of the computational model is shown in Fig.~\ref{fig:multidisp diagram uav}.
For the laser-beam model, we apply the Beers--Lambert law for optical power transfer~\cite{kim2001comparison} to compute the power received by the aircraft as
\begin{equation}
P_{\text{rec}} = \eta P_{\text{tra}}, \quad  \eta = e^{-\sigma R},
\end{equation}
where $P_{\text{rec}}$ and $P_{\text{tra}}$ denote the power received and transmitted, respectively;
$\sigma$ is the atmospheric extinction coefficient which differs across weather conditions;
and $R$ is the distance between the aircraft and the power source.
In the aerodynamic model, inputs such as flight velocity, flight altitude, and the total aircraft weight are considered. This model is responsible for computing the total drag force exerted on the wing to sustain steady-state flight. Additionally, it determines the lift coefficient and drag coefficient by employing a linear model and a quadratic polar model, respectively, derived from airfoil data.
The weights model estimates the weight of each aircraft component using the statistical equations from~\cite{raymer2012aircraft} and outputs the total aircraft weight.
The performance model takes inputs such as flight velocity and total drag force to compute the total energy stored during the mission. This computation is based on the following equations:
\begin{equation}
\begin{aligned}
    & n  = \sqrt{\left( \frac{v}{rg}\right)^2+1}, \\
    & P_{\text{req}}  = \frac{D}{n}, \\
    & E  = (P_{\text{rec}} - P_{\text{req}})t,
\end{aligned}
\end{equation}
where $n$ represents the turn load, $r$ is the radius of the mission, $g$ is the gravitational acceleration, $D$ denotes the total drag force, $P_{\text{req}}$ is the required power to maintain steady flight, and $t$ represents the mission time.

On this test problem, following our proposed method, sparsity ratios of the uncertain inputs are measured, which are shown in Tab.~\ref{tab: sparsity_uav}.
\begin{table}[]
\caption{Sparsity ratio of the uncertain inputs in the UAV model}
\centering
\begin{tabular}{c | c} 
 Uncertain inputs & Sparsity ratio \\ 
 \hline
$v$ & 92\% \\
$h$ & 96\%   \\
$W_p$ & 4\%  \\
$\sigma$ & 2\%  \\
\end{tabular}
\label{tab: sparsity_uav} 
\end{table}
Using the sparsity ratios, two sparse uncertain inputs are identified: $W_p$ and $\sigma$, as both of them have a sparsity ratio less than 5\%. Combining this information with our understanding of the computational model, we constructed the partially tensor-structured quadrature rule as:
\begin{equation}
\label{eqn: opt_ten_struc}
    \boldsymbol{u}  = \boldsymbol{u}^{n_1}_{\{h, v\}} \times  \boldsymbol{u}^{n_2}_{\{W_p\}} \times  \boldsymbol{u}^{n_3}_{\{\sigma\}}.
\end{equation}

\begin{figure}%
    \centering
    \subfloat[\centering Convergence plots without AMTC]{{\includegraphics[width=7cm]{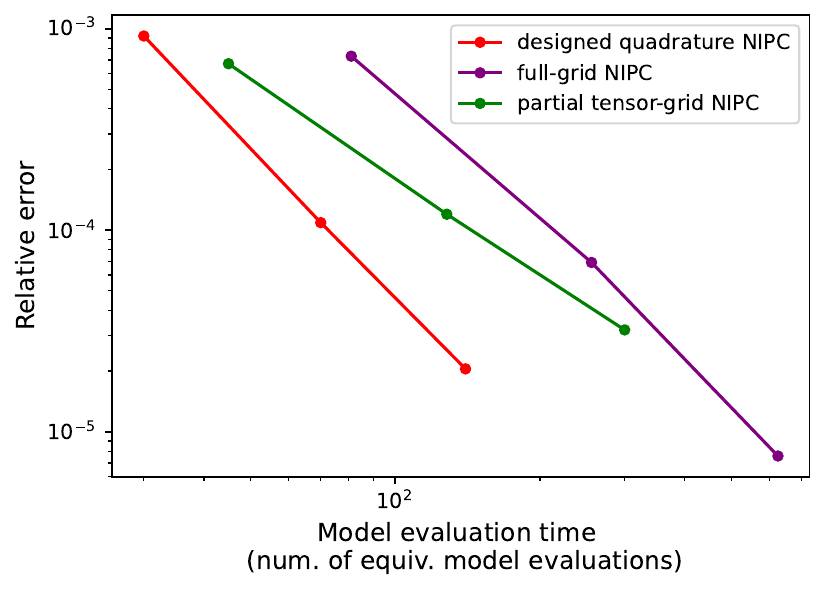} }}
    \qquad
    \subfloat[\centering Convergence plost with AMTC]{{\includegraphics[width=7cm]{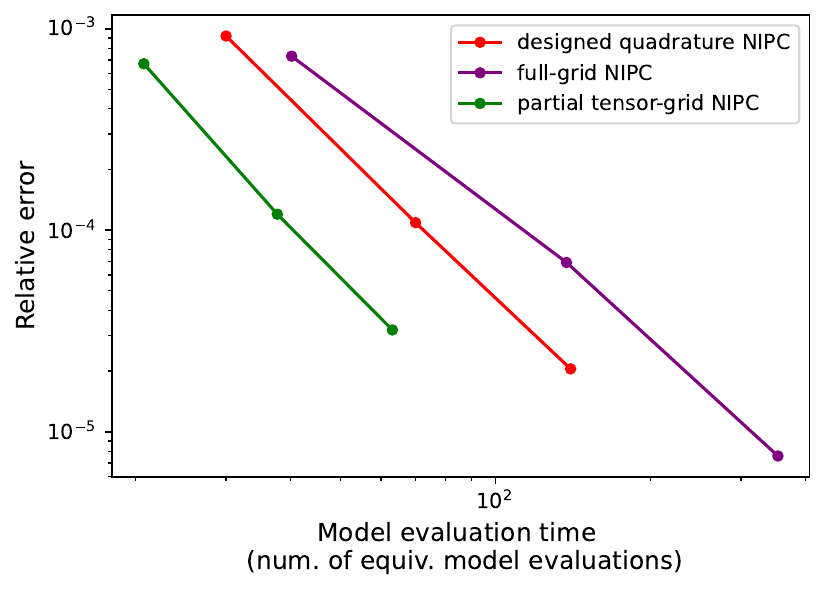}}}
    \caption{Convergence plots with and without AMTC for the UAV problem}%
    \label{fig:uav_result}%
\end{figure}

Our proposed method is compared with the designed quadrature method and the full-grid Gauss quadrature method, all of which are used with the NIPC method to solve the UQ problem. 
The UQ convergence plots, with and without using AMTC, are shown in Fig.~\ref{fig:uav_result}. 
From the results, we observe that the partially tensor-structured quadrature rule is outperformed by the designed quadrature method when AMTC is not used to accelerate the model evaluations. 
This finding aligns with our theoretical analysis, as the non-tensorial quadrature rule requires fewer quadrature points than the tensor-structured quadrature rules to achieve the same level of accuracy.
However, when we utilize AMTC to accelerate the model evaluations, the advantages of the structured quadrature rules become evident. 
Both the model evaluations associated with the full-grid quadrature rule and the partially tensor-structured quadrature rule are significantly accelerated by using the AMTC method. 
In this scenario, the partially tensor-structured quadrature rule achieves the best performance, with more than 50\% reduction in evaluation costs than the other two methods. This is because, our method constructed a tailored tensorial structure that makes the most effective use of the inherent sparsity within the computational graph of this multidisciplinary system, resulting in the lowest model evaluation costs under the same level of accuracy.

\subsection{Air-taxi trajectory design multidisciplinary model}
The second test problem is a 6D UQ problem adapted from an air-taxi trajectory optimization scenario. 
This computational model calculates the average ground-level sound pressure level during the aircraft's flight along a specific trajectory. 
The UQ problem aims to determine the expected output under uncertainties in the control inputs and parameters of the acoustic model.
Three uncertain inputs are associated with the initial conditions of the trajectory, 
and the other three uncertain inputs are associated with the parameters in the acoustic model. 
The distributions of the uncertain inputs can be found in Tab.~\ref{tab:airtaxi_uncertain_inputs}.
Notably, the control inputs used in this problem are generated from solving a deterministic trajectory optimization problem, which aims to minimize the total energy expended throughout the trajectory. Details of the computational model along with the trajectory optimization formulation can be found in~\cite{orndorff2023air}.
\begin{table}[h!]
    \centering
    \begin{tabular}{c | c }
         Uncertain inputs & Distributions \\
         \hline
         \text {Initial velocity $v_0$ ($m/s$)} & {$\mathcal{U}(30, 35)$} \\
         \text {Initial flight path angel $\gamma_0$ ($\deg$)} & {$\mathcal{U}(0, 2)$} \\
         \text {Initial altitude $h_0$ ($m$)} & {$\mathcal{U}(0, 20)$} \\
       $\beta_1$ & $\mathcal{N}(0.0209, 0.001)$ \\
       $\beta_2 $ & $\mathcal{N}(18.2429, 0.5)$ \\
       $\beta_3 $ & $\mathcal{N}(6.729, 0.2)$ \\
    \end{tabular}
    \caption{Uncertain inputs' distributions for the air-taxi problem}\label{tab:airtaxi_uncertain_inputs}
\end{table}

\begin{figure}[hbt!]
\centering
  \includegraphics[width= 16 cm]{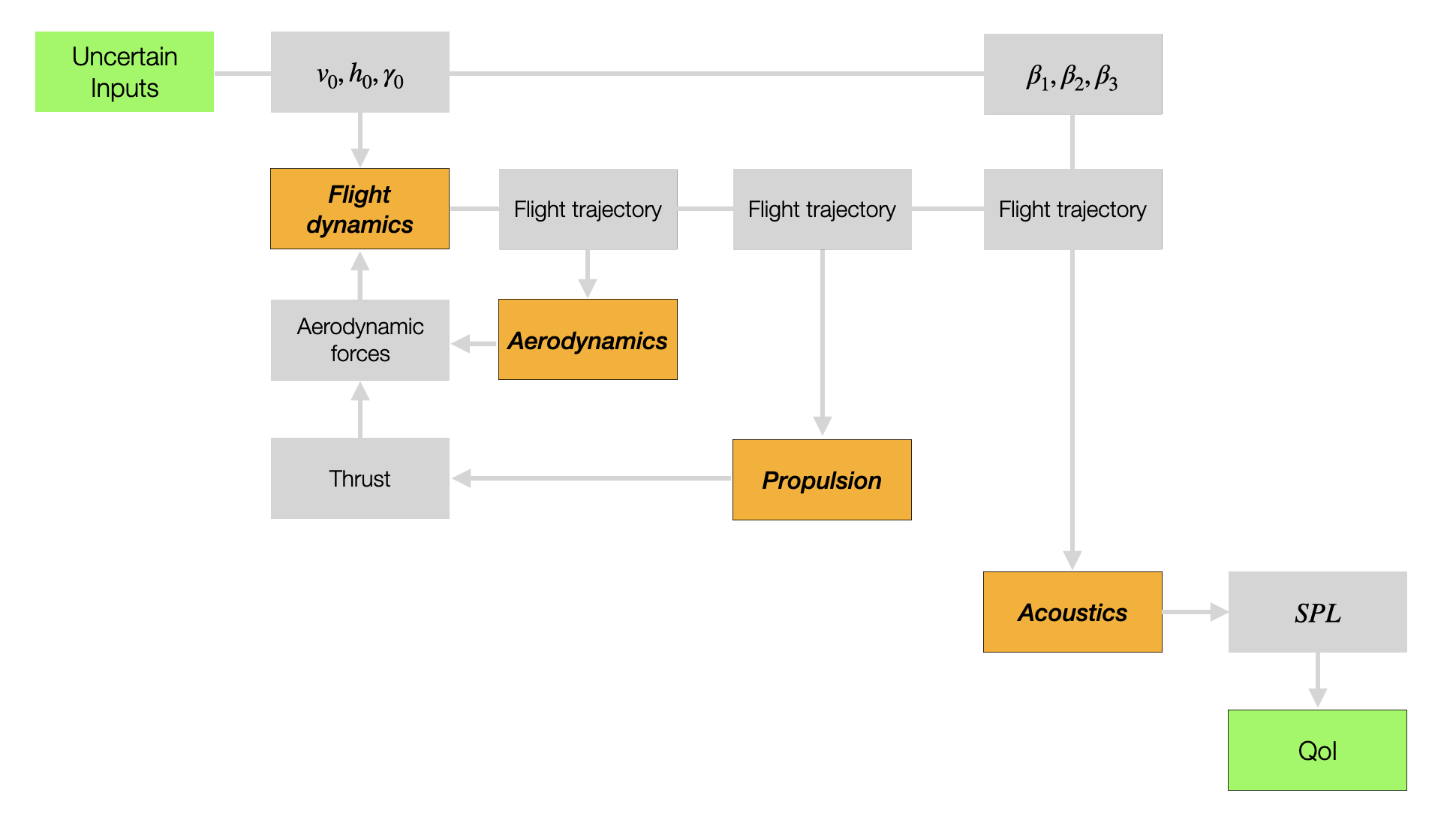}
\caption{Multidisciplinary structure of the air-taxi model}
\label{fig:airtaxi_multi}
\end{figure}

In this problem, the measured sparsity ratios are shown in Tab.~\ref{tab: airtaxi_sparsity}. 
Based on this information, three uncertain inputs are identified as sparse uncertain inputs which are the three uncertain inputs associated with the acoustic model. 
This finding aligns with the multidisciplinary structure of the computational model, depicted in Fig.~\ref{fig:airtaxi_multi}.
From this figure, we observe that the computational model comprises four disciplines: flight dynamics, aerodynamics, propulsion, and acoustics, with two-way coupling among the first three disciplines.
In this case, the coupled disciplines functions as a nonlinear operation and is solved iteratively. 
Due to its dominant role in computational costs, the three uncertain inputs associated with the nonlinear solver are the dense uncertain inputs in this computational model.
Conversely, the parameter uncertain inputs are the sparse uncertain inputs as they do not affect the upstream nonlinear solver in this computational model.
Based on this reasoning, we choose the partially tensor-structured option as 
\begin{equation}
    \boldsymbol{u}  = \boldsymbol{u}^{n_1}_{\{v_0, h_0, \gamma_0\}} \times  \boldsymbol{u}^{n_2}_{\{\beta_1, \beta_2, \beta_3\}},
\end{equation}
in which we form a tensor structure between the space of dense uncertain inputs and the space of sparse uncertain inputs.
\begin{table}[]
\caption{Sparsity ratio of the uncertain inputs in the air-taxi model}
\centering
\begin{tabular}{c | c} 
 Uncertain inputs & Sparsity ratio \\ 
 \hline
$v_{0}$ & 99\% \\
$h_{0}$ & 99\%   \\
$\gamma_{0}$ & 99\%   \\
$\beta_1$ & 2\%  \\
$\beta_2$ & 2\%  \\
$\beta_3$ & 2\% \\
\end{tabular}
\label{tab: airtaxi_sparsity} 
\end{table}

\begin{figure}%
    \centering
    \subfloat[\centering Convergence plot without AMTC]{{\includegraphics[width=7cm]{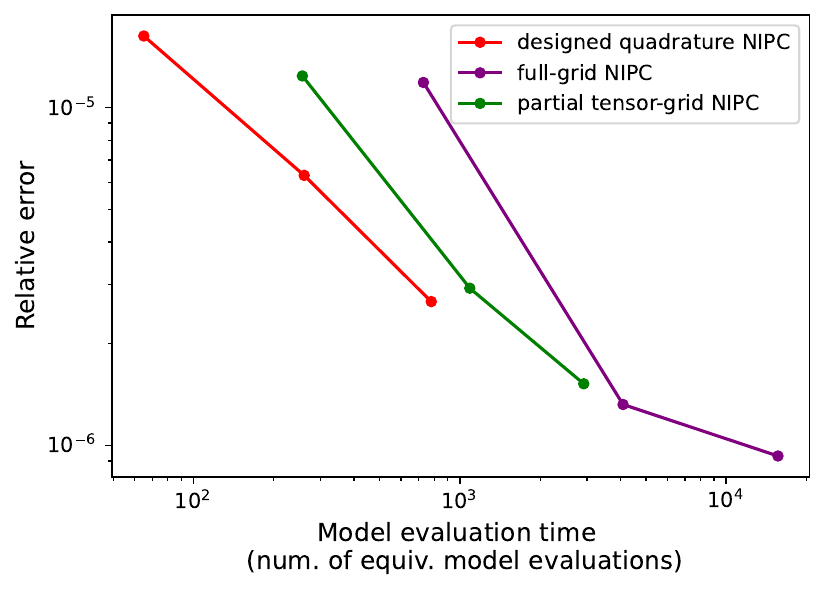} }}
    \qquad
    \subfloat[\centering Convergence plot with AMTC]{{\includegraphics[width=7cm]{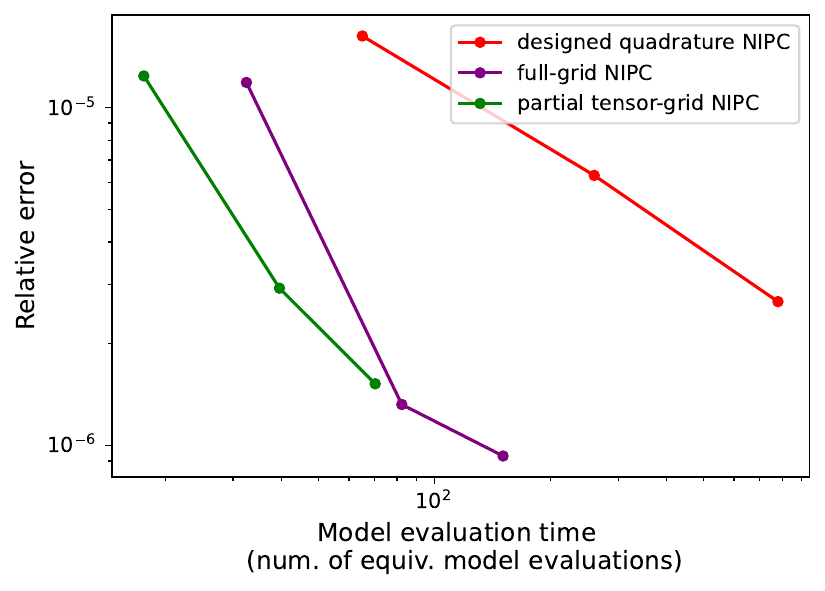}}}
    \caption{Convergence plots with and without AMTC for the airtaxi problem}%
    \label{fig:airtaxi_result}%
\end{figure}

The UQ convergence plots with and without using AMTC are shown in Fig.~\ref{fig:airtaxi_result}. Similar to the first test problem, the partially tensor-structured quadrature rule is outperformed by the designed quadrature method when AMTC is not used. 
However, it becomes the most efficient method when using AMTC to accelerate the model evaluations. 
The reduction in computational costs is more than 40\% in most cases.  This can be explained by the fact that the tailored tensorial structure option takes full advantage of the inherent sparsity of the computational graph, resulting in the most efficient performance with the AMTC method.
However, we also observe that when the desired relative error is exceptionally low (1e-6 and lower), the performance of the partially tensor-grid quadrature rule deteriorates on this test problem. 
This phenomenon can be attributed to the fact that the designed quadrature rule method can achieve similar performance to the Gauss quadrature rule while requiring a significantly smaller number of model evaluations. 
However, as the dimension of the optimization problem increases (as we have more optimization variables and more constraints), it becomes more challenging to solve the optimization problem to a tight tolerance using this method. Consequently, the designed quadrature method may struggle to achieve the same level of accuracy as the Gauss quadrature rule when an extremely high level of accuracy is required.

\section{Conclusion}
\label{Sec: Conclusion}

In this paper, we introduced a novel approach for generating partially tensor-structured quadrature rules tailored for utilization with the graph-accelerated NIPC method to tackle UQ problems.
Our method entails initially choosing an appropriate tensorial structure option for the quadrature rule through an analysis of the computational graph.
Subsequently, the quadrature rule possessing the chosen tensorial structure is generated using the designed quadrature method.
This method has been tested on two UQ problems: one with four uncertain inputs and the other with six uncertain inputs, both involving multidisciplinary systems and are derived from aircraft design scenarios.
The numerical results show that our method generates a new, partially tensor-structured quadrature rule for both problems. 
Notably, this new rule exhibits superior performance compared to both the existing designed quadrature method and the full-grid Gauss quadrature method when used with the graph-accelerated NIPC method.

One limitation of this proposed method is its reliance on the inherent sparsity of the computational graph, which restricts the generation of new quadrature rules for every problem.
However, our method improves the efficiency of the graph-accelerated NIPC method, particularly for medium-dimensional UQ problems (typically with fewer than 10 dimensions) involving multidisciplinary systems.
This is achieved by generating partially tensor-structured rules to more efficiently leverage the inherent sparsity within the computational graph.
Another limitation lies in the potential performance decline of our method for high-dimensional UQ problems (ten dimensions or higher), where alternative approaches such as Monte Carlo methods may outperform it. 
A promising direction for future research involves integrating this method with dimension-reduction techniques to address higher-dimensional UQ problems. Another interesting research direction would be to develop non-tensor quadrature methods that are invariant under axis permutations, ensuring that flipping the axis vectors results in the same quadrature rule.
\section*{Acknowledgments}

The material presented in this paper is, in part, based upon work supported by  DARPA under grant No.~D23AP00028-00.


\bibliography{sample}
\end{document}